\documentclass[epj]{svjour}
\usepackage{graphics}
\usepackage{graphicx}
\usepackage{epsfig}
\usepackage{amsmath}
\usepackage{amssymb}
\usepackage{amsfonts}
\begin{document}
\title{Crime and punishment: the economic burden of impunity}
\author{Mirta B. Gordon\inst{1} \and J. R. Iglesias\inst{2} \and
Viktoriya Semeshenko\inst{1} \and J.-P. Nadal\inst{3}}                     
%
%
\institute{Laboratoire TIMC-IMAG (UMR 5525), University of Grenoble I, Domaine de La Merci - Jean Roget, F-38706 La Tronche, France \and Instituto de F\'{\i}sica and Faculdade de Ci\^encias Econ\'omicas, Universidade Federal do Rio Grande do Sul, 91501-970 Porto Alegre, Brazil \and Laboratoire de Physique Statistique (UMR8550), Ecole Normale Sup\'erieure, 24, rue Lhomond,
F-75231 Paris Cedex 05 and Centre d'Analyse et de Math\'ematique Sociales (CAMS) (UMR8557) Ecole des Hautes Etudes en Sciences Sociales,
54, Bd. Raspail, F - 75270 Paris Cedex 06, France
}
\date{Received: date / Revised version: date}
%
\abstract{
Crime is an economically important activity, sometimes called the
``industry of crime". It may represent a mechanism of wealth distribution
but also a social and economic charge because of the cost
of the law enforcement system. Sometimes it may be less costly
for the society to allow for some level of criminality. A drawback of
such policy may lead to a high increase of criminal activity that
may become hard to reduce. We investigate the level of law enforcement
required to keep crime within acceptable limits and show that a sharp
phase transition is observed as a function of the probability of
punishment. We also analyze the growth of the economy, the inequality in
the wealth distribution (the Gini coefficient) and other relevant
quantities under different scenarios of criminal activity and probability of apprehension.
} 
\maketitle

\newpage

\section{Introduction}
Crime is a human activity probably older than the crudely called
``oldest profession''. Criminal activity may have many different
causes: envy, like in the case of Cain and Abel~\cite{genesis},
jealousy like in the opera Carmen, or financial gain like Jacob
cheating Esau with the lentil pottage (and his father with the lamb
skin) to obtain the birthright \cite{genesis2}. This is to say that
crime can have many different causes, some of them ``passional",
sometimes ``logical" or ``rational" \cite{VolterraConsulting03}.

All along history, organized societies have tried to prevent and to
deter criminality through some kind of punishment. In all the societies
and all the times punishment has been in some way proportional to the
gravity of the offense. Methods ranged from the {\em lex talionis} ``an
eye for an eye'', to fines, imprisonment, and the death penalty. Most
of the literature on crime considered criminals as deviant individuals.
Usual explanations of why people offend use concepts like insanity,
depravity, anomia, etc. In 18th Century England criminals were
massively ``exported" to Australia because it was thought that the
criminal condition was hereditary and incurable: incapacitation was the
solution.

In any case the increase of criminal activity in different countries
have led some sectors of the population, as well as politicians, to ask
for harder penalties. The understated idea is that a hard sentence,
besides incapacitation of convicted criminals, would have a deterrent
effect on other possible offenders and would also prevent recidivism.
Yet, the deterrent effect of punishment is a polemic subject and law
experts diverge with respect to whether offenders should be
rehabilitated or simply punished (see for example the recent public
discussion in France \cite{Salas06}). In many countries, incapacitation
is the main reason for imprisonment of criminals. For instance the
Argentinean constitution states that prisons are not there for the
punishment of the inmates but for the security of the society:
criminals are isolated, not punished \cite{argent}.

Although the idea that the decision of committing a crime results from
a trade off between the expected profit and the risk of punishment
dates back to the eighteen and nineteenth centuries, it is only
recently that crime modeling emerged as a field worth of being
investigated.

In a review paper, Alfred Blumstein \cite{Blumstein02} traces back the
recent interest in crime modeling to 1966, when the USA President's
Commission on Law Enforcement and Administration of Justice created a
Task Force on Science and Technology. Composed mainly by engineers and
scientists, its aim was to introduce simulation modeling of the
American criminal justice system. The model allowed to evaluate the
resource requirements and costs associated to a criminal case, from
arrest to release, by considering the flow through the justice system.
For example, it estimated the opportunity of incarceration of convicted
criminals and the length of the incarceration time.
In his concluding remarks, Blumstein states that: ``{\em We are still
not fully clear on the degree to which the deterrent and incapacitation
effects of incarceration are greater than any criminalization effects
of the incarceration, and who in particular can be expected to have their
(tendency to)\footnote{We added these words in parenthesis}
crime reduced and who might be made worse by the punishment}".

In a now classical article G. Becker \cite{Becker68} presents for the
first time an economic analysis of costs and benefits of crime, with
the aim of developing optimal policies to combat illegal behavior.
Considering the social loss from offenses, which depends on their
number and on the produced harm, the cost of aprehension and conviction
and the probability of punishment per offense, the model tries to
determine how many offenses should be permitted and how many offenders
should go unpunished, through minimization of the social loss function.

Using a similar point of view, Ehrlich \cite{Ehrlich75,Ehrlich96}
develops an economic theory to explain participation in illegitimate
activities. He assumes that a person's decision to participate in an
illegal activity is motivated by the relation between cost and gain, or
risks and benefits, arising from such activity. The model seemed to
provide strong empirical evidence of the deterrent effectiveness of
sanctions. However, according to Blumstein \cite{Blumstein02} the
results ``{\em ... were sufficiently complex that the U.S. Justice
Department called on the National Academy of Sciences to convene a
panel to assess the validity of the Erlich results. The report of that
panel highlighted the sensitivity of these econometric models to
details of the model specification, to the particular time series of
the data used, and to the sensitivity of the instrumental variable used
for identification, and so called into question the validity of the
results}". In fact, the issue of what constitutes an optimal crime
control policy is still controversial.

Another controversial subject is the relationship between raising of
income inequality and victimization. Becker's economic model of crime
would suggest that as income distribution becomes wider, the richer
become increasingly attractive targets to the poorer. There are many
reasons why this hypothesis may not be correct. For example, Deutsch et
al. \cite{DeutschSpiegelTempleman92} consider the impact of wealth
distribution on crime frequency and, contrary to the general consensus
in the literature, conclude that variations of the wealth differences
between ``rich" and ``poor" do not explain variations in the rates of
crime. Bourguignon et al \cite{BourguignonNunezSanchez02} based on data
from seven Colombian cities conclude that in crime modeling the average
income of the population determines the expected gain, but that
potential criminals belong to the segment of the population whose
income is below $80 \%$ of the mean. More recently, Dahlberg and
Gustavsson \cite{DahlbergGustavsson05} pointed out that in crime
statistics one should distinguish between permanent and transitory
incomes. Disentangling these income components based on tax reports in
Sweden, they find that an increase in inequality in permanent income (measured through the
variance of the distribution) yields a positive and
significant effect on crime rates, while an increase in the inequality
due to a transitory income has no significant effect. Levitt \cite
{Levitt99} concludes from empirical data that, probably because rich
people engage in behavior that reduces their victimization, the trends
between 1970 and 1990 is that property crime victimization has become
increasingly concentrated on the poor.

As pointed out by many authors, the fact that police reduces crime is
far from being demonstrated. Realizing that (at least in the U.S.A.)
the number of police officers increases mostly in election years,
Levitt \cite{Levitt97} has studied the correlation between these
variations (uncorrelated to crime) and variations in crime reduction.
He finds that increases in the size of police forces substantially
reduce violent crime, but have a smaller impact on property crime.
Moreover, the social benefit of reducing crime is not larger than the
cost of hiring additional police. Similar conclusions have been drawn
by Freeman \cite{FreemanRB96}, who estimates that the overall cost of
crime in the US is of the order of 4 percent of the GDP, 2 per cent
lost to crime and 2 percent spent on controlling crime. This amounts to
an average of about $54,000$ dollars/year for each of the 5 million or
so men incarcerated, put on probation or paroled.

In most models, punishment of crime has two distinct aspects: on one
side there is the frequency at which illegal actions are punished
(which corresponds to the punishment probability in the models), and on
the other, the severity of the punishment. In a review paper, Eide \cite
{Eide99} comments that although many empirical studies conclude that
the probability of punishment has a preventive effect on crime, the
results are ambiguous.

More recent reviews of the research literature consider the factors
influencing crime trends \cite{VolterraMarris00} and present some
recent modeling of crime and offending in England and Wales \cite
{VolterraConsulting03}. They both note the above mentioned difficulties
in estimating the parameters of the models and the cost of crime.

Most of the models in the literature follow Becker's economic approach.
Ehrlich \cite{Ehrlich96} presents a market model of crime assuming that
individuals decisions are  ``rational": a person commits an offense if
his expected utility exceeds the utility he could get with legal
activities. At the ``equilibrium" between the supply of crimes and the
``demand'' (or tolerance) to crime --- reflected by the expenditures for
protection and law enforcement -- neither criminals, private
individuals nor government can expect to improve their benefits by
changing their behaviors. In particular, the model is based on standard
assumptions in economic models, with ``well behaved" monotonic supply
and demand curves, which cannot explain situations where social
interactions are important \cite{NaPhGoVa05,GoNaPhSe07}.  Indeed,
Glaeser et al \cite{GlaeserSacerdoteScheinkman96} attribute to social
interactions the large variance in crime on different cities of the US.

Another kind of models \cite{CampbellOrmerod00,VolterraConsulting03}
treat criminality as an epidemics problem, which spreads over the
population due to contact of would-be criminals with ``true" criminals
(who have already committed crime). This kind of models incorporates
effects due to social interactions, which introduce large
nonlinearities in the level of crime associated to different
combinations of the parameters. These may explain the wide differences
reported in the empirical literature.

In this paper we focus on economic crimes, where the criminal agents try
to obtain an economic advantage by means of the accomplished felony. No
physical aggression or death of the victim will be considered, and on
the side of punishment we consider the standard of most developed
civilized countries, i.e. fines and imprisonment. We assume that
\cite{web_dando} {\em ``most criminal acts are not undertaken by deviant
psychopathic individuals, but are more likely to be carried out by
ordinary people reacting to a particular situation with a unique
economic, social, environmental, cultural, spatial and temporal context.
It is these reactionary responses to the opportunities for crime which
attract more and more people to become involved in criminal
activities rather than entrenched delinquency''}.

We simulate a population of heterogeneous individuals. They earn
different wages, have different tastes for criminality, and modify
their behaviors according to the risk of punishment. We assume that
both the probability and the severity of the punishment increase with
the magnitude of the crime. We are interested in the consequences of
the punishment policy on the costs of crime, in the wealth distribution
consistent with different levels of criminality, in the economic growth
of the society as a function of time, and in the (possibly bad)
consequences of allowing for some criminal activity in order to
minimize the cost of the law enforcement apparatus. In section
\ref{sec:Description of the model} we describe the model, in section
\ref{sec:Monthly dynamics} we explain its dynamics. We present
the results of the simulations in \ref{sec:Simulation results} and leave the
conclusions to section \ref{sec:Discussion and conclusion}

\section{Description of the model}
\label{sec:Description of the model}
We consider a model of society with a constant number of agents,
$N$, who perceive a periodic (monthly) positive income $W_i$ (that we
call wage), spend part of it each month and earns the rest. Wages remain
constant during the simulated period. In our simulations, the wages
distribution has a finite support: $W_i \in [W_{min},W_{max}]$.
In this utopian society there
is not unemployment, and it is assumed that the minimum wage is
enough to provide for the minimum needs of each agent, i.e. a person
perceiving the minimum wage will expend it completely within the month.
These wages and possibly the booties of successfully realized crimes
constitute the only income source of the individuals.
Besides the living expenses, capitals may decrease due to plunders
and to the taxes or fines related to conviction, as explained later.

We assume  that each agent has an inclination to abide by the
law, that is represented by a honesty index $H_i$ ($i \in [1,N]$)
which, at the beginning of the simulations ranges between a minimum and
a maximum: $H_i \in [H_{min},H_{max}]$. This inclination --- that may
be psychological, ethical, or reflect educational level and/or
socio-economical environment --- is not an intrinsic characteristic of
the individuals. It changes from month to month according to the risk
of apprehension upon performing a crime.

In our simulations, the individual decision of committing a crime
depends (not exclusively) on both the honesty index and the monthly
income, which are initially drawn at random from
distributions $p_H$ and $p_W$ respectively, without any correlation
among them. This is justified by the lack of empirical evidence that
the poorer are more or less law abiders than the rich.

The honesty index of all the individuals but the criminal increases by
a small amount each time a crime is punished. Otherwise it decreases,
but at a different rate for the criminal than for the rest of the
population. The honesty index is not affected by the importance of the
punishment. This assumption is an extreme simplification of the
observation \cite{Blumstein02} that the crime rate is more sensitive to
the risk of apprehension than to the severity of punishment. We have
studied different distributions $p_H$ and $p_W$ and different
treatments of the honesty index. The latter differ in the way we treat
the lower bound of the distribution $p_H$, namely $H_{min}$. Hereafter we
describe in details the simplest case: we consider that $H_{min}$ is an
{\em absolute} minimum of the honesty. $H_{min}$ corresponds thus to
the honesty index of the most recalcitrant offenders. But we also add
the hypothesis (that is certainly controversial) that for those agents
with this lowest honesty there is no possible redemption, i.e. when the
honesty index of an agent decreases down to $H_{min}$ it remains there
for ever. We leave to forthcoming discussions the possible variations
of this scheme.

As a consequence of this treatment, there are intrinsic criminals
(those with $H_i=H_{min}$). Indeed, a finite fraction, $N_C(0)$, of
intrinsically criminal agents is assumed from the very beginning
of the simulation, according with the idea that there always have
been and will be a finite number of not redemptible criminals in a
society. All the $N-N_C(0)$ other agents are ``susceptible'': their
degree of honesty $H_i(0)$, drawn at random from the probability
distribution $p_H$, satisfies $H_{min} < H_i(0) \le
H_{max}$.

Individuals have also ``social'' connections: encounters between
criminals and victims may only occur between socially linked
individuals. These connections are not meant to represent social
closeness but rather the fact that those individuals share common
daily trajectories or live in close neighborhoods, or meet each other
just by chance. For simplicity, here we consider the latter situation:
every individual may be connected to every other individual.

Notice that the nature of social interactions in our model is very
different from the mimetic interactions considered by Glaeser et al
\cite{GlaeserSacerdoteScheinkman96}, or the social pressure introduced by
Campbell et al. \cite{CampbellOrmerod00}. We study just the case of
individual crimes. No illicit criminal associations (``maffia'') nor
collective victims (like in bank assaults) are going to be considered.

The characteristics of the simulated population evolve on time. We
assume that every month there is a number of possible criminal attempts.
This number depends on the honesty index of the population as is
explained later. Among these attempts, some are successful, i.e. the
criminal spoils his victim of a (random) fraction of his earnings.

In contrast with most models, crimes are punished with a probability
that depends on the magnitude of the loot. When punished, the criminal
returns part of the booty to the victim, pays a fine that may be
considered as a contribution to the public enforcement system and goes
to jail for a number of months that also depends on the stolen amount.
Maintaining each criminal in prison bears a fixed cost per month to the
society, that we evaluate.

In our simulations we study the month to month evolution of different
quantities that characterize the system, and how these depend on
the probability of punishment.

\section{Monthly dynamics}
\label{sec:Monthly dynamics}
Starting from initial conditions of honesty, wages and earnings
described in section \ref{sec:initial}, we simulate the model for a
fixed number of months. Within each month, there is a random number of
criminal attempts correlated with the honesty level of the population
(section \ref{sec:attempts}). However, not all the attempts end up with
a crime: the potential criminal has to satisfy some reasonable
conditions that we explicit in section \ref{sec:criminals}. If there is
crime, there is a transfer of the stolen value from the victim to the
criminal, as detailed in section \ref{sec:crime}. Some of the offenders
are detected and punished (section \ref{sec:punishment}): part of the
stolen amount is then returned to the victim and a fine or tax is
levied from the criminal's earnings. In turn, the honesty of the
population evolves after each crime, according to the success or
failure in crime repression (section \ref{sec:honesty}).

We keep track of the individuals' wealth, taking into account the
monthly incomes, the living expenses, the plunders, and the cost of
imprisonment, as detailed in section \ref{sec:earnings}.

In the following paragraphs we describe in details the dynamics of the
model.

\subsection{Initial conditions}
\label{sec:initial}
The individuals have initial honesty indexes $H_i(0)$ and wages $W_i$
drawn with uncorrelated probability density functions (pdf) $p_H$ and
$p_W$ respectively. We assume that the
majority of the population is honest and also that the majority earn
low salaries, i.e. $p_H$ has a maximum at large $H$ while $p_W$ has its
maximum at small $W$. The results described below have been obtained
considering triangular distributions, because they are the simplest way
of introducing the desired individual inhomogeneities. Thus, $p_W$ is a
triangular distribution of wages, with $W_{min}=1$ and $W_{max}=100$
(in some arbitrary monetary unit), with its maximum at $W_{min}$ and a
mean value $\overline W = W_{min}+(W_{max}-W_{min})/3$. As already
explained, there is an initial number of ``intrinsic" criminals $N_C(0)$
 drawn at random, with $H_i=H_{min}$. In our simulations this number
has been set to $5 \%$ of the population ($N_C(0)=0.05 N$). The honesty
index of the remaining (susceptible) individuals is triangular, from
$H_{min}=0$ up to $H_{max}=100$, with the maximum at $H_{max}$.

Individuals' initial endowments are arbitrarily set to five months
wages: $K_i(0)=5 \, W_i$. This initial amount controls the time needed
for the dynamics to fully develop. Smaller initial endowments result in
transients dominated by the size of the possible loots, because these
cannot exceed the victims' capitals.

\subsection{Attempts}
\label{sec:attempts}
The number of criminal attempts each month $m$ is $A(m)=[{\cal A}(m)]$
where $[ \dots]$ represents the integer part, and
\begin{equation}
\label{eq:attempts}
{\cal A}(m)=\left\{
\begin{array}{lll}
1 + r(m) \, c_A \, N + & \frac{H_{max}(m)-\overline H(m)}{\overline H(m)}
\ N_C(m) & \\ & {\rm if}  \overline H(m) \neq H_{min}&, \\
  \\
1 + r(m) \, c_A \, N + & 2 \, N_C(m) & \\ & {\rm if}  \overline H(m) = H_{min}&.
\end{array}
\right.
\end{equation}
$r(m) \in [0,1]$ is a random number drawn afresh each month, $0 < c_A <
1$ is a coefficient (in the simulations $c_A=0.01$) and $N_C(m)$ are
the number of intrinsic offenders (who have $H_i=0$) at month $m$. With
these considerations there is always a ground level of criminal
attempts given by the term proportional to $N$. Criminality increases
with the number of intrinsic offenders and with a decreasing average
honesty. When the latter equals $H_{min}$ all the population has the
minimum of honesty, i.e. everybody is an ``intrinsic'' offender. In this
case, to avoid the divergence in the first of equations (\ref
{eq:attempts}), the prefactor of $N_C$ was set arbitrarily to $2$.

At each criminal attempt a potential offender is selected as explained
in subsection \ref{sec:criminals} and the number of remaining attempts
for the month is decreased by one, even if the crime is aborted.

\subsection{Criminals}
\label{sec:criminals}
We consider that the possible criminals have low honesty indexes, and
that it is more likely that they have low wages. Then, at each attempt,
a {\em potential} criminal $k$ is drawn at random among the population
of free (not imprisoned) individuals (we discuss in
\ref{sec:punishment} when and how long punished offenders go in jail).
We also select at random an upper bound $H_U$ to the honesty index in
the interval $[H_{min},H_{max}]$. If
\begin{equation}
\label{eq:conditionH}
H_k < H_U,
\end{equation}
then the offense takes place with a probability:
\begin{equation}
\label{eq:conditionW}
p_k=e^{-W_k/\overline W},
\end{equation}
that is higher the lower the potential criminal wages. Thus, we select a
random number in $[0,1)$. If it is larger than $p_k$,
the criminal attempt is aborted.

\subsection{Crime}
\label{sec:crime}
If crime is actually committed, a {\em victim} $v$ among the
social neighbors (not in jail) of $k$ is selected at random.
We do not take into account the wealth of the victim, which may be
an important incentive for the criminal's decision. Thus, our treatment
may be adequate for minor larcenies, like stealing in public
transportation. In the crimes considered in this paper, the victim is
robbed of a random amount $S$ that cannot be larger than his actual
capital. In the simulation, this amount is proportional to the
victim's wage:
\begin{equation}
\label{eq:stolen}
S = r_S \, c_S \, W_v,
\end{equation}
where $r_S \in [0,1]$ is a random number drawn afresh at each
attempt and $c_S > 0$ is a coefficient (here we use $c_S =10$,
implying that robbery may concern amounts up to 10 times the victim's
wage). If the value given by (\ref{eq:stolen}) is
larger than the victim's capital $K_v$, we set $S=K_v$.
Correspondingly, the victim's capital is decreased by the stolen
amount, to become $K_v-S$ while the criminal's capital is
in turn increased to $K_k+S$.

\subsection{Punishment}
\label{sec:punishment}
The criminal may be catched and punished with probability
\begin{equation}
\label{eq:punishment}
\pi(S)=\frac{p_1}{1 + \frac{p_1-p_0}{p_0} e^{-\frac{S}{\overline
K(m)}}},
\end{equation}
a monotonic function that starts at its minimal value $p_0$ (for $S=0$
), increases almost linearly with $S$ for values of $S$ smaller than
$\approx \overline K(m)$, the population's average wealth, and gets
close to the asymptotic value $p_1$ for large booties.

A punished criminal is imprisoned for a number of months $1 +
[S/\overline W]$ (the square brackets mean integer part), {\em i.e.}
proportional to the stolen amount. During this time he does not earn
his monthly wage and cannot be selected as a potential offender nor as
a victim.

Beyond incarceration, the convicted criminal suffers a financial loss.
He is deprived of an amount $(f_R + f_D) S$, larger than the
loot but that cannot exceed his total wealth (we do not allow for
negative capitals). It is worth to remark that the financial loss is not
limited to reimburse the stolen capital. A monetary punishment (a fine)
is inflicted to the offender together with the time in prison (that represents,
in addition to be incarcerated, also a monetary loss). The total amount deduced from the
offender's capital is composed of a fraction $f_R S$ (with $f_R<1$), that is
returned to the victim, and an amount, $f_D S$ (or $K_k - f_R S$ if  $f_D S > K_k$),
considered as a duty. Cumulated duties or ``taxes'' constitute the total
income of fines. In this simulation $f_R=0.75$ and
$f_D=0.45$, meaning that both criminals and victims contribute to
taxes, since the victims only recover a fraction $1-f_R$ of the stolen
amount. However, with the assumed values of $f_R$ and $f_D$, $f_R + f_D = 1.2$, implying that the
criminals also afford part of the costs.

\subsection{Honesty dynamics}
\label{sec:honesty}
We assume that punishment has a dissuasive effect on the population,
although not necessarily on the convict. Thus, whenever a criminal
is convicted, the honesty index of all the population {\em but the
criminal}, is increased by a fixed amount $\delta H$. Convicted
criminals do not change their honesty level. On the contrary, when the
crime is not punished, the entire population but the criminal decrease
their honesty index by $\delta H$, while the criminal decreases his by $2
\delta H$: unpunished criminals become even less honest.

In the present simulations we do not allow $H_i$ to become negative.
Moreover, the lower bound of the distribution is absorbing. Thus,
individuals reaching $H_i=H_{min}$ have their honesty index freezed,
and henceforth are considered as ``intrinsic" criminals.

A dynamics with a non-absorbing bound for the honesty dynamics gives
similar results to those presented here. A variant that we did not
implement yet is to modify the honesty level of the criminal
proportionally to the importance of the loot.

\subsection{Monthly earnings and costs}
\label{sec:earnings}
At the end of each month $m$, after the $A(m)$ criminal attempts
are completed, the individuals' cumulated capitals
$K_i(m)$ are updated: the total capital of each agent, $K_i(m)$, is
increased with his salary and decreased with his monthly expenses.
Notice that the criminals' and the victims' capitals have been
further modified during the month, according to the results of the
criminal attempts.

We assume that individuals need an amount $W_{min} + f
\times (W_i-W_{min})$ with $f < 1$ to cover their monthly expenses.
Thus, individuals with higher wages spend a proportional part of
their income in addition to the minimum wage $W_{min}$. This is a
simplifying assumption which may be questionable since the
richer the individuals, the smaller the fraction of income they
need for living but, on the other hand, rich people spent more in luxury goods. Assuming a more involved model for the expenses
would modify the monthly wealth distribution, making it more
unequal, but we do not expect that the qualitative results of our
simulations would be modified.

In order to quantify the expenditure of conviction and imprisonment, we
assume that the monthly cost of maintaining a criminal in jail is equal
to the minimal wage. This is clearly a too simple hypothesis, that does
not take into account the fixed costs of maintaining the public
enforcement against crime. It is just a mean of assessing some social
cost proportional to the criminal activity. The cumulated taxes,
obtained through punishment of convicted criminals are thus decreased
by an amount $W_{min}$ per month for each convict in jail.

At the end of each month, the time that inmates have to remain in
prison (excluding the criminals convicted during that month) is
decreased by one; those having completed their arrest punishment
are freed.

\section{Simulation results}
\label{sec:Simulation results}
Starting with the initial conditions, there are $A(m)$ criminal
attempts each month $m$ as described in the subsection \ref{sec:attempts}. As a consequence of the criminal
activity, during the month the capitals of criminals and victims are modified, as
well as the honesty index of the population. With the above
dynamics, the earnings and the honesty distributions are shifted and
distorted. The system may even converge to a population where all
the initially susceptible individuals end up as intrinsically
criminal.

We simulated systems of $N=1 \, 000$ agents for a period of $240$
months, under different distributions of honesty and wages. Here we
report results corresponding to the triangular distributions
described in section \ref{sec:Description of the model}: wages have
a linearly decreasing distribution in $[W_{min}, W_{max}]$ with
$W_{min}=1$ and $W_{max}=100$: there are more individuals with low
wages than with high ones. The honesty distribution is also
triangular, but increasing: starting with $p_H(H_{min})=0$, it
increases linearly reaching its maximum at $p_H(H_{max})$, which
corresponds to assuming that there are more honest than dishonest
individuals.

\subsection{Evolution of criminality with time}
\begin{figure*}
        \begin{center}
        \includegraphics[width=8cm]{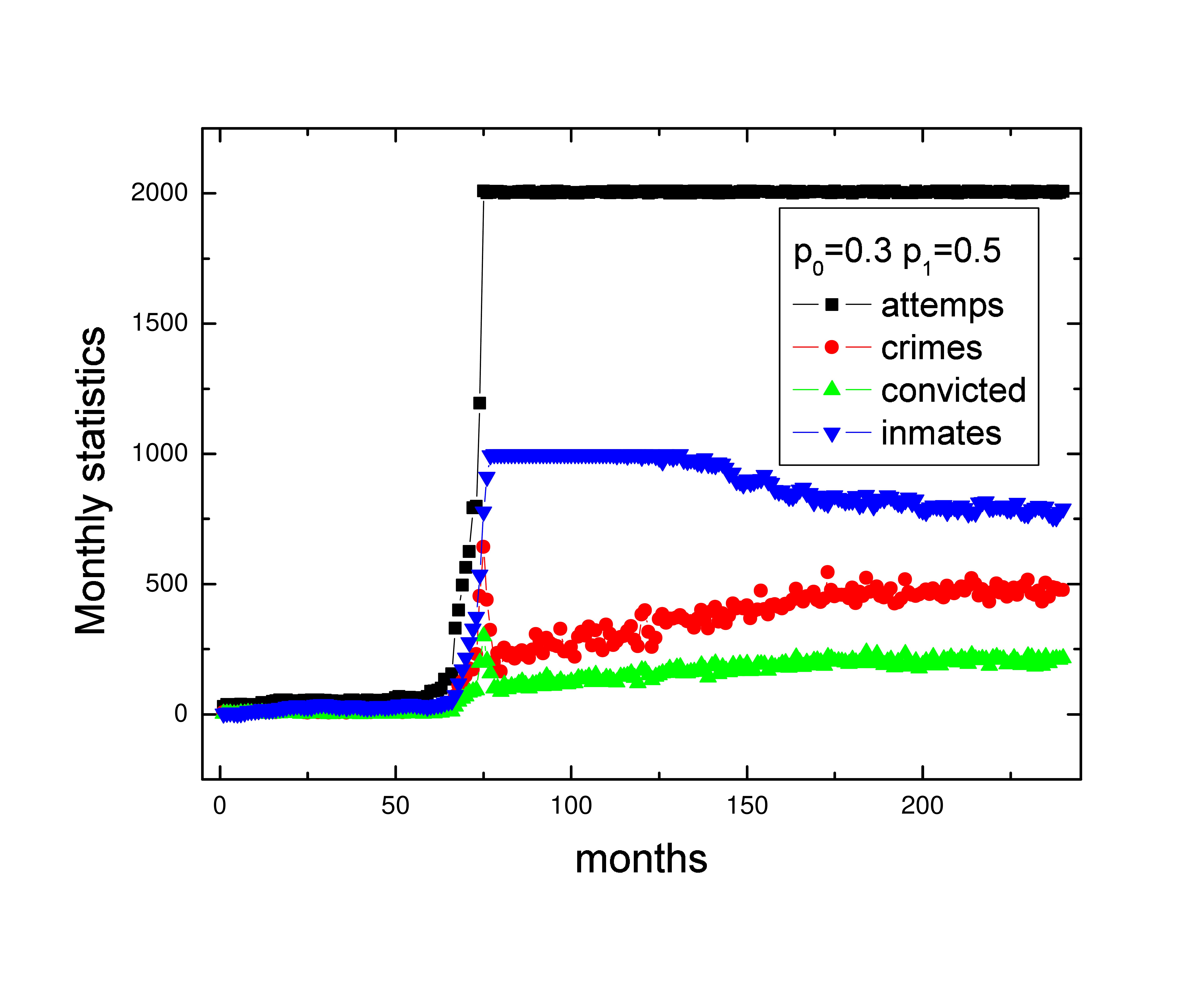}
        \includegraphics[width=8cm]{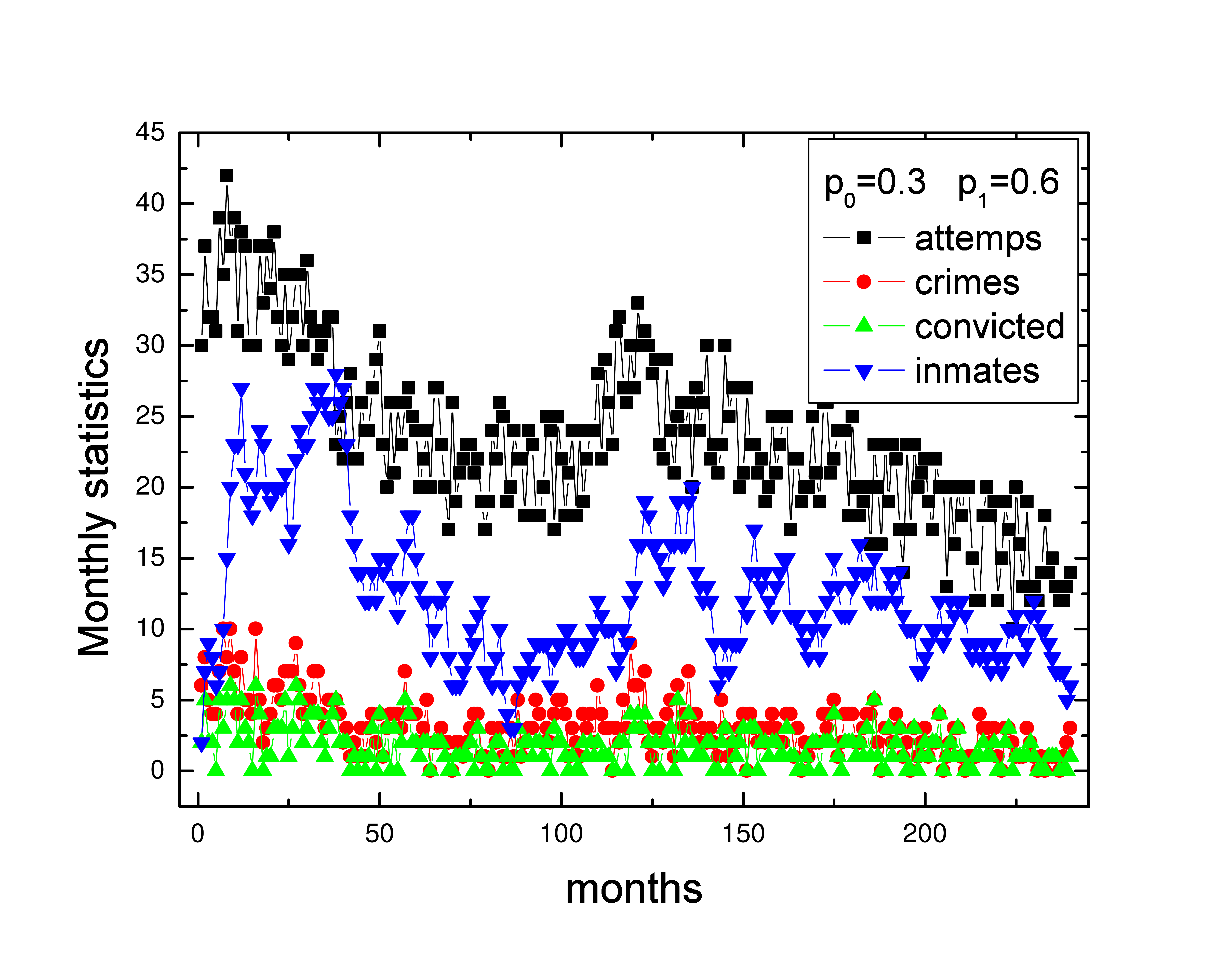}
        \end{center}
		\caption[Criminals monthly statistics] {Statistics of
criminality per month. Attempts: number of criminal attempts in the
month. Crimes: number of crimes committed in the month. Convicted:
number of criminals punished in the month. Inmates: number of criminals in
jail. The difference between the left (a) and the right (b) panels is the
maximum punishment probability $p_1$, $p_1=0.5$ in the left panel and
$p_1=0.6$ in the right panel. Therefore, a small change in the
probability of punishment induce a enormous change in the criminality
(see the difference in scale in the ordinate axis).}
\label{fig:Monthstat}
\end{figure*}

\begin{figure*}
        \begin{center}
        \includegraphics[width=8cm]{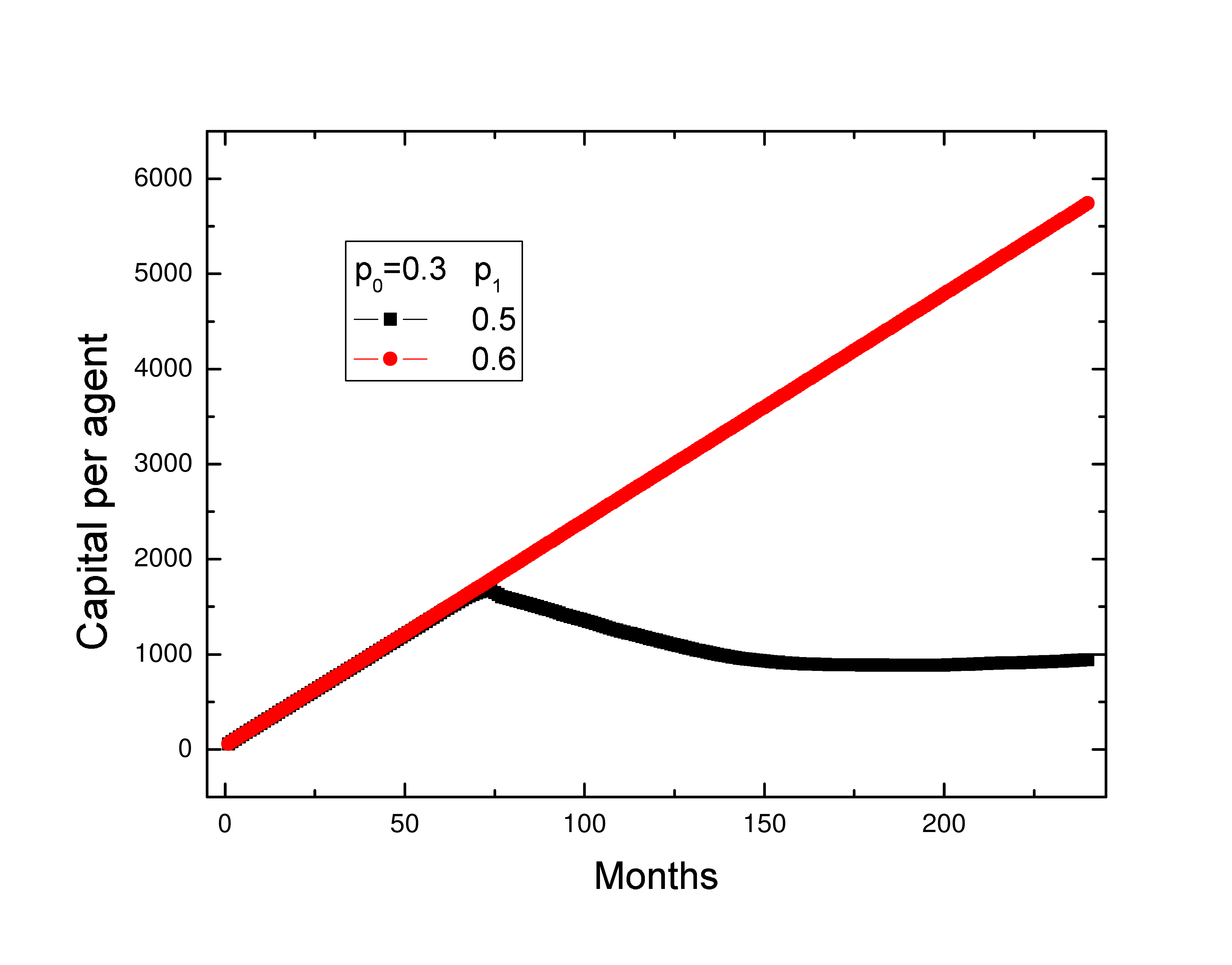}
        \includegraphics[width=8cm]{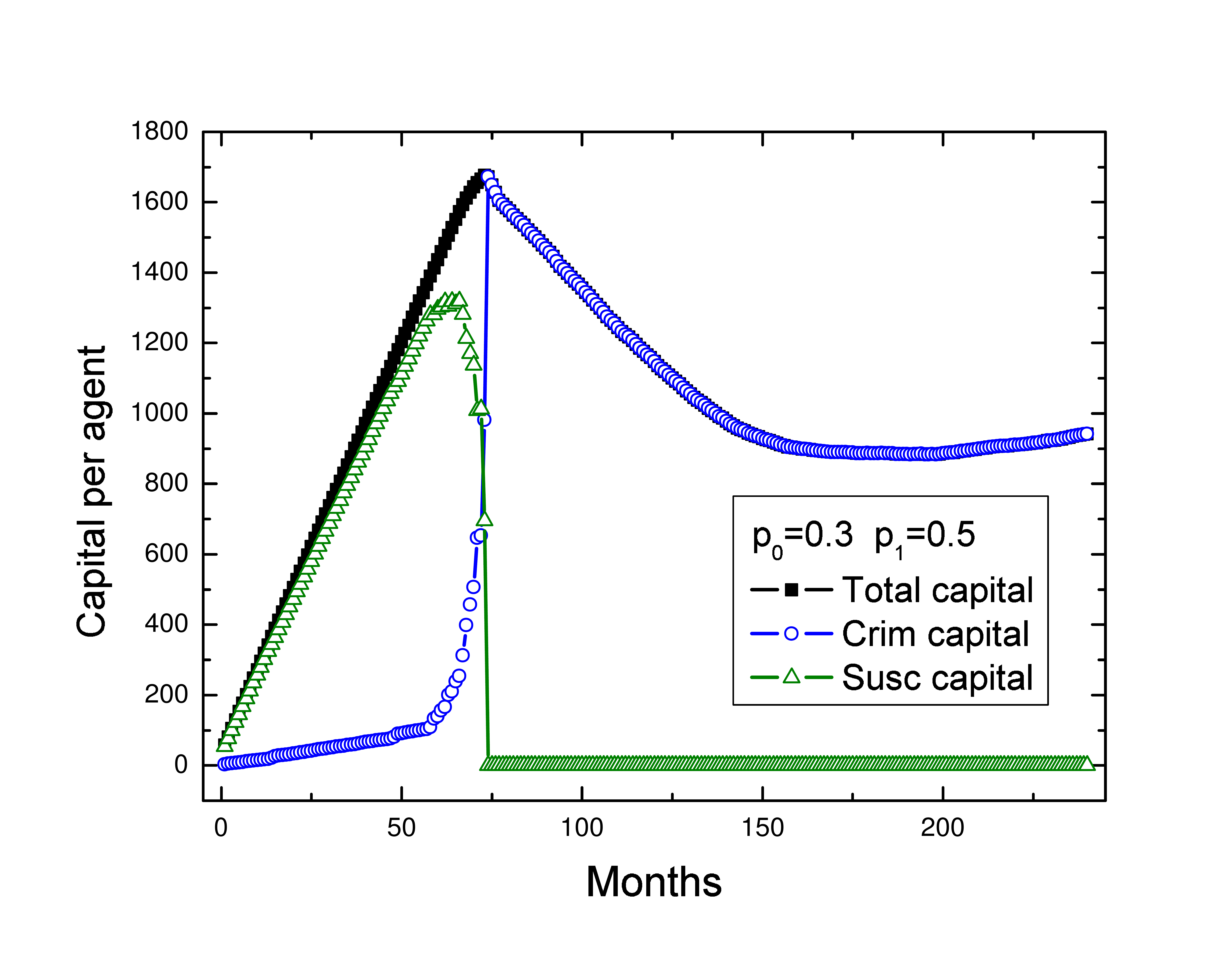}
        \includegraphics[width=8cm]{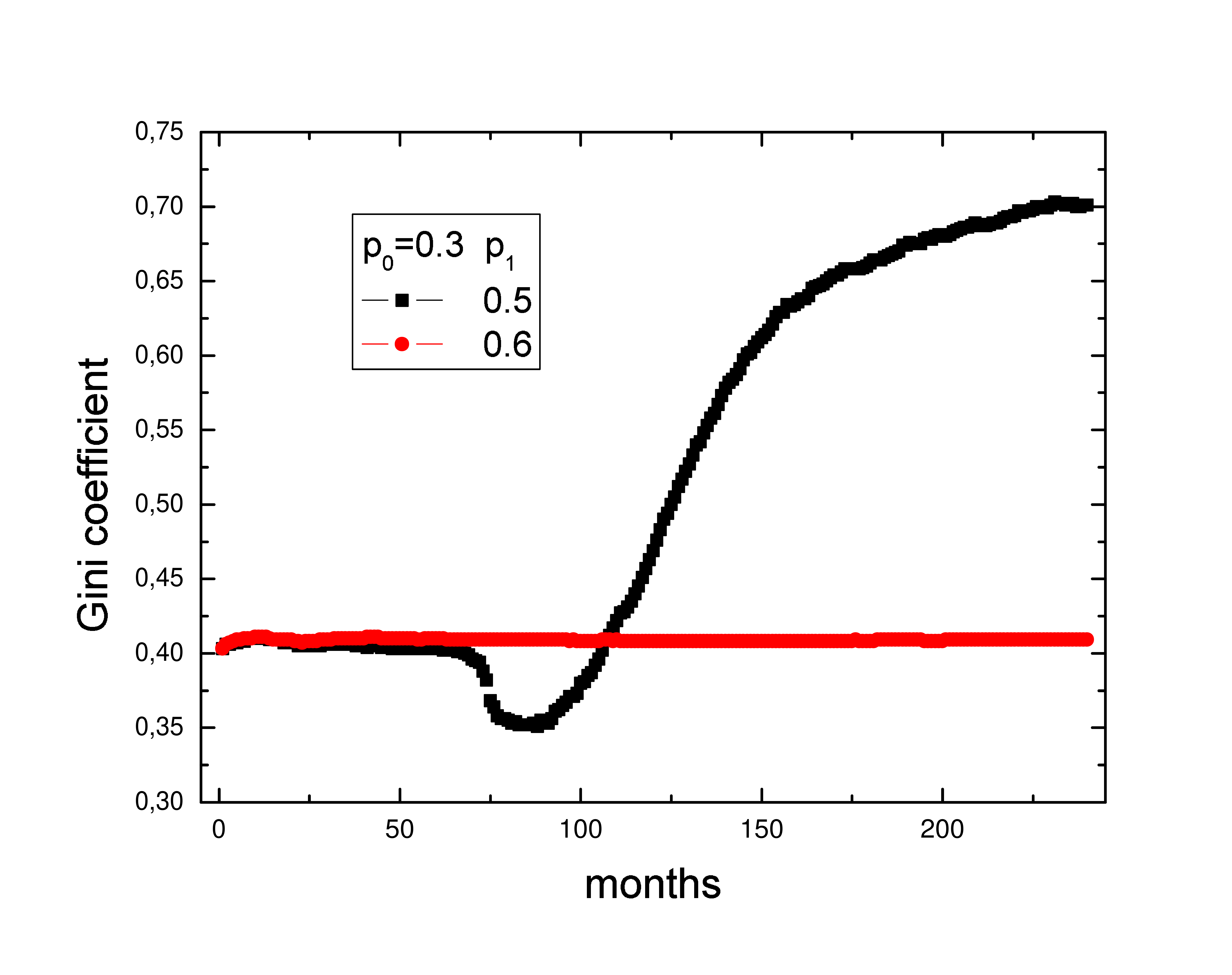}
        \end{center}
		\caption[Capitals monthly statistics] {Capital and
Gini monthly statistics: panel (a) shows
the cumulated capital per agent as a function of time
(the black curve corresponds to $p_1=0.5$ and the red one to $p_1=0.6$),
panel (b) shows the participation in the total capital of intrinsic
criminals (blue line) and susceptible ones (green line) just
for the case of high criminality, $p_1=0.5$; panel (c) shows the
Gini coefficient for two values of $p_1$, as in panel (a).}
\label{fig:MonthKapitalGini}
\end{figure*}

The system's evolution is very sensitive to the punishment
probability. Generally, the first months the rate of crime is low
because the number of intrinsic criminals is small and the stolen
amounts cannot be larger than the initial endowments. But after some
months (the number depends on the initial conditions and on the
values of $p_0$ and $p_1$) the crime rate increases to a stationary
value.

Figure \ref{fig:Monthstat} shows a typical monthly evolution of the
number of attempts, the number of crimes, the number of convicted
criminals (which are the criminals punished in the corresponding
month) and of inmates. The left figure corresponds to relatively low
values of both $p_0$ (the probability of punishing small offenses)
and $p_1$ (the probability of punishing large offenses) in (\ref
{eq:punishment}). Beyond about $75$ months the number of attempts
increases through an avalanche to reach its saturation value, while
the number of inmates grows to reach almost all the population,
producing a drop in crime. This drop is not due to a deterrent effect
of punishment but
rather to the absence of possible criminals (they are all in jail).
Eventually the system evolves to a smaller number of inmates, but
the population is essentially composed of individuals with the smallest
honesty index: the punishment rate is not enough to keep
criminality below any acceptable level. The figure on the right (see
the change in the ordinate scale) corresponds to a slightly higher
value of $p_1$, showing the impressive impact of increasing the
probability of punishment of big offenses. All the quantities
(attempts, crimes, etc.) present a dramatic decrease with respect to
the values in the left hand side figure. Notice that the fluctuations on
the reported quantities are of the same order of magnitude in both
figures: they are due to the probabilistic nature of the quantities
involved (see equations (\ref{eq:attempts}) to (\ref{eq:punishment}) ).

Correspondingly, the earned capital per individual (figure
\ref{fig:MonthKapitalGini} upper left) increases almost linearly
with time when crime is limited. In contrast, in the highly criminal
society ($p_1=0.5$) it begins to decrease as soon as the regime of
high criminality is reached, mainly because most criminals are
convicted and do not receive their wages, but also because, since the
number of punished crimes is also high, the amount retained
in the form of taxes is discounted from the total wealth of the
society. Moreover, the high cost of imprisonment, proportional to
the total number of inmates, drains also part of the capital and
contributes to decrease the capital per capita. This is
illustrated on figure \ref{fig:MonthKapitalGini} upper right, where
the total capital per agent is decomposed into the capital hold by
the intrinsic criminals and the part hold by the susceptible
population (with honesty index $H_i > 0$). When the system reaches
the high criminality regime the latter drops to zero because there
are no more susceptible individuals. Clearly, when the
level of punishment is not high enough to guarantee an effective
control of criminality, the cost of the repressive system is very
high. It is remarkable that a very small increase of the upper value
$p_1$ in (\ref{eq:punishment}) is enough for a complete change in the
scenario.

The lower figure \ref{fig:MonthKapitalGini} presents the evolution
of the Gini index of the population, defined by:
\begin{equation}
\label{eq:gini}
G=\frac{\sum_{j=1}^{N-1}\sum_{i=j+1}^{N}|K_i-K_j|}{(N-1)\sum_i^N
K_i}.
\end{equation}
The Gini index spans in the interval $[0,1]$ and measures the inequality in the
wealth distribution of the population. Its minimal value, $G=0$,
corresponds to a perfectly equalitarian society. The Gini index of the
initial endowment, distributed according with a triangular probability
density, is $G(0)=0.4$. Due to the dynamics, when crime level is
moderate ($p_1=0.6$) $G$ is seen to slightly increase. However, in
the high criminality regime ($p_1=0.5$) it oscillates, and when the high
crime rate sets in, it first plummets down because successful
criminals, mostly individuals with small incomes according to the
probability of crime (\ref{eq:conditionW}), increase their wealths
at the victims' expense. As a result, the wealth distribution
becomes more evenly distributed. However, on the long run, the Gini
index increases dramatically. This is so because if everybody is a
lawbreaker (lowest honesty index) criminals and victims are the same,
just one stealing the other. So when one agent is in the victim role he
becomes poor because of the robbery, and when he behaves as a crook he
also finishes poor (generally), mainly because he pays taxes and also
does not earn his wage when in jail. So, just a few agents are able to
hold large capitals thanks to crime, increasing inequality in this
society.

In fact, when $p_0$ is smaller than a critical value of the order of
$0.5$, on increasing $p_1$, the system presents an abrupt
transition between a high crime --- low honesty population to a low
crime --- high honesty one. This transition, apparent on all the
quantities, as may be seen below on figure \ref{fig:Versus_p_1},
corresponds to a swing of the system between a regime of high
criminality to one where the criminality level is moderately low.

In the high criminality side, cumulated earnings are small, taxes
are high and the Gini index is large. Conversely, on the low
criminality side, {\em i.e.} for sufficiently high $p_1$, the
cumulated wealth increases monthly according to the earned wages,
and the Gini index reflects the distribution of the latter. We will show in Figures
\ref{fig:WealthHistograms}, below, typical histograms of wealth
distribution at both sides of the transition, as well as the initial
distribution.

\subsection{Changing the punishment probability}
\label{sec:changingprob}

In the previous section we have discussed the time evolution of
criminality. Let's now consider the final state of the society (after
240 months) as a function of the probability of punishment. In order to
make comparable experiments we have studied societies with the same
initial conditions subjected to different levels of punishment. Notice
that these levels are constant during the simulated 240 months.

\begin{figure*}
        \begin{center}
        \includegraphics[width=8cm]{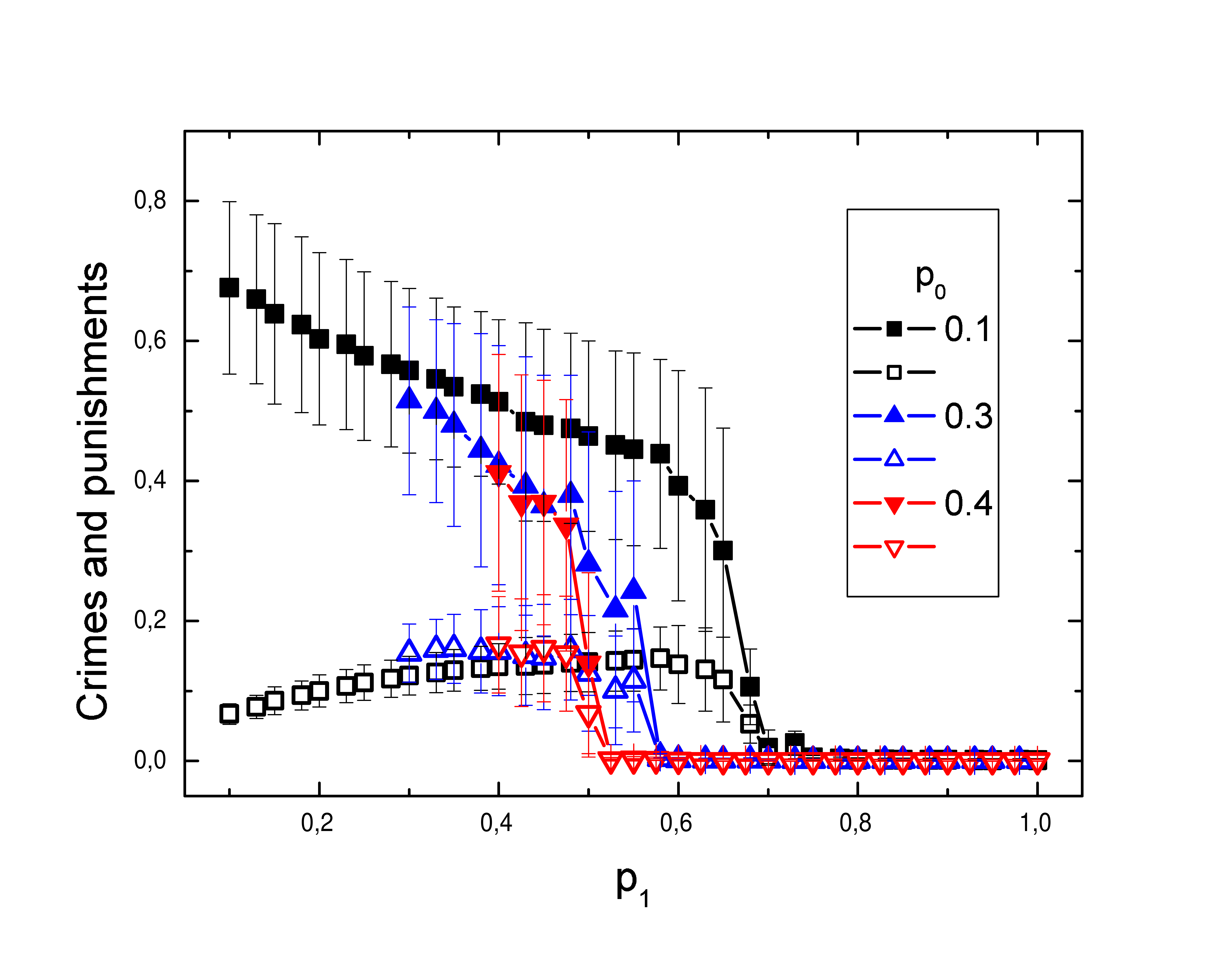}
        \includegraphics[width=8cm]{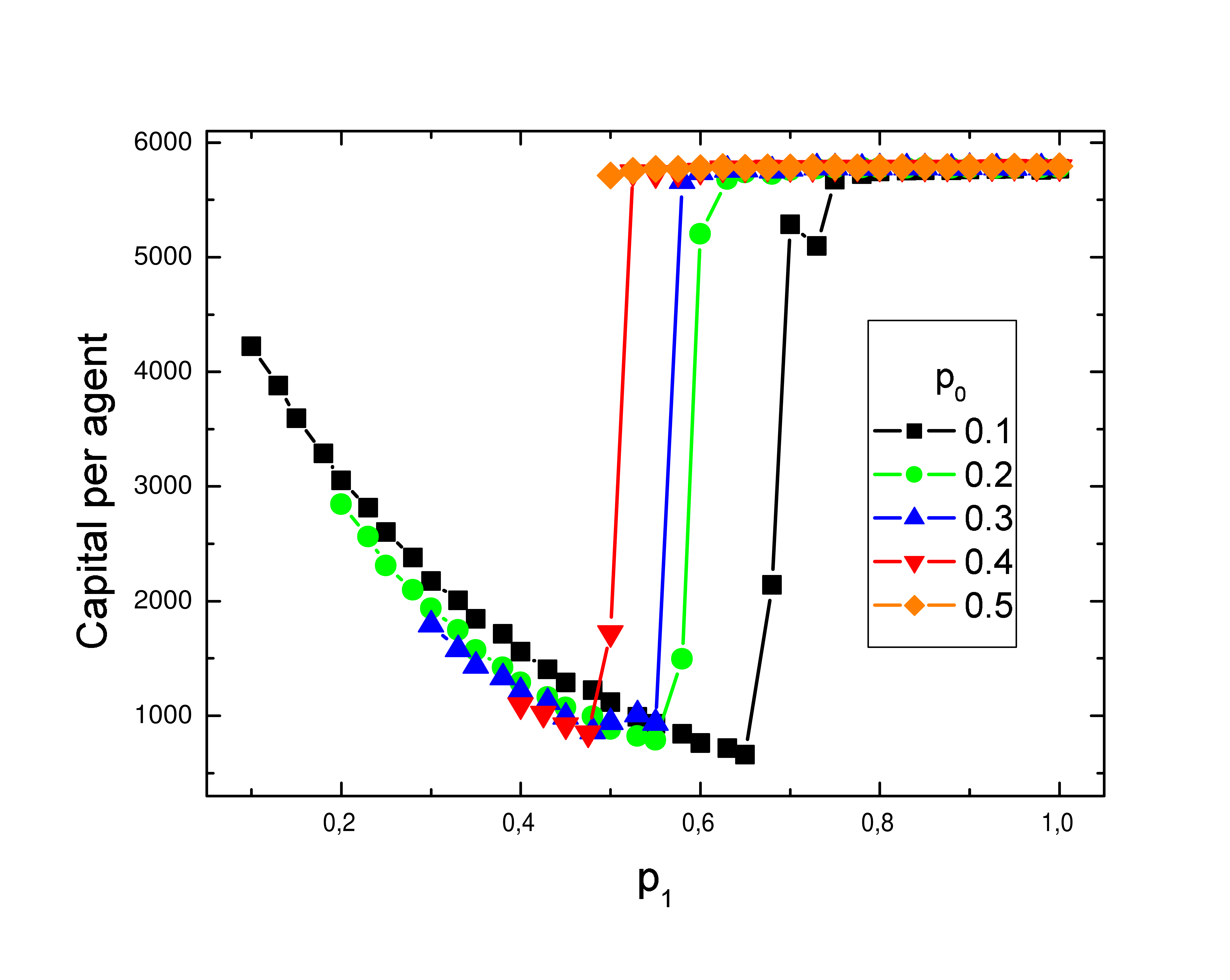}
        \includegraphics[width=8cm]{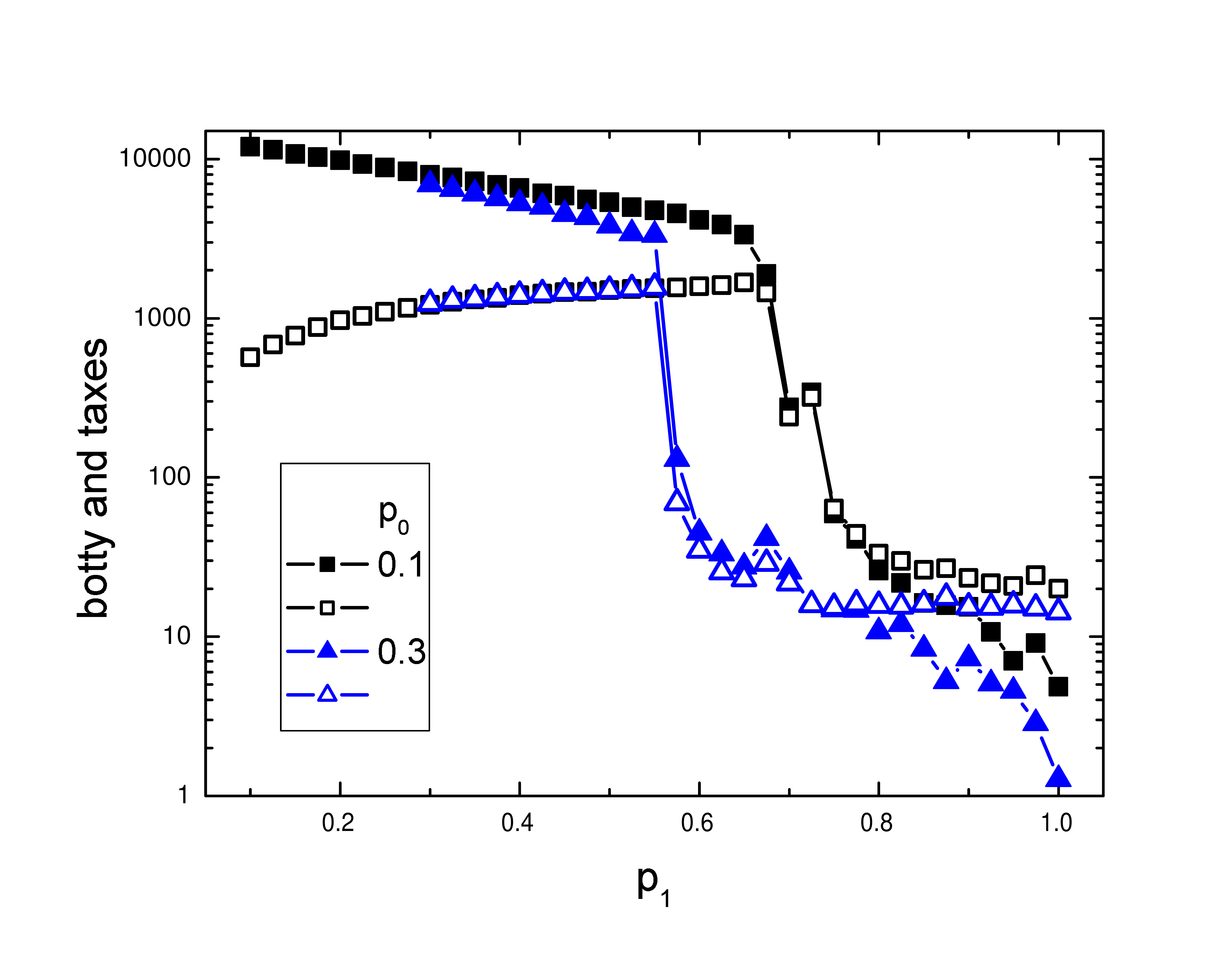}
        \includegraphics[width=8cm]{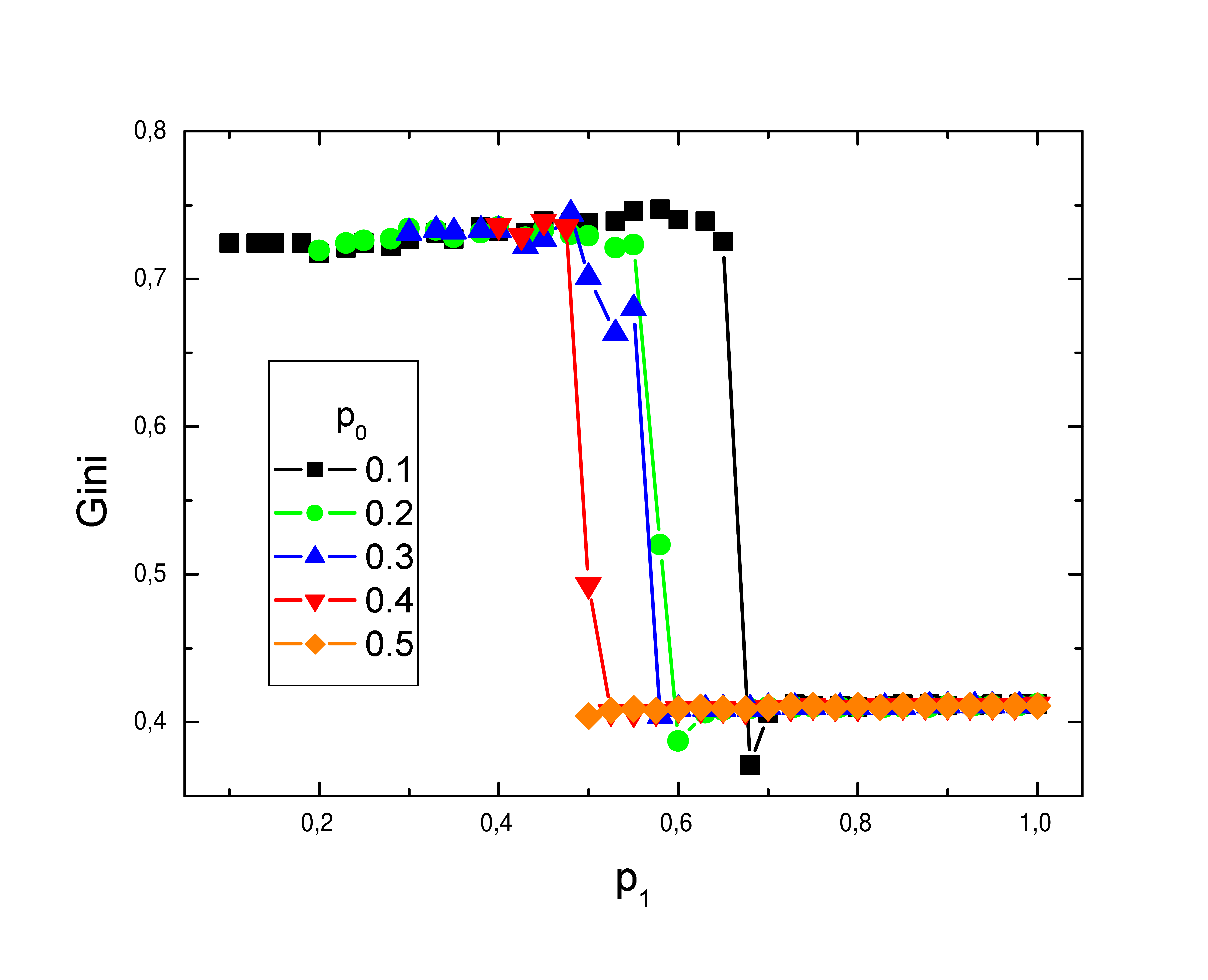}
        \end{center}
        \caption[Values vs $p_1$] {Panel (a): average number of crimes
and convicted criminals over the simulated period. Panel (b): last month
capital per capita of the population. Panel (c): average loot (full
symbols) and taxes (empty symbols). Panel (d): last month Gini index.}
\label{fig:Versus_p_1}
\end{figure*}

We consider different values of $p_0$ smaller than a critical value of
the order of $0.5$, and we study the variation of several social
indicators as a function of $p_1$. We observe that the system ends up with
either a high crime-low honesty population (for $p_1$
lower than a critical value) or a low crime-high honesty one.
This transition, apparent on all the quantities, as may be seen on
figures \ref{fig:Versus_p_1}, corresponds to a swing of the system
between a regime of high criminality to one of moderate criminality.

\begin{figure*}
        \begin{center}
        \includegraphics[width=8cm]{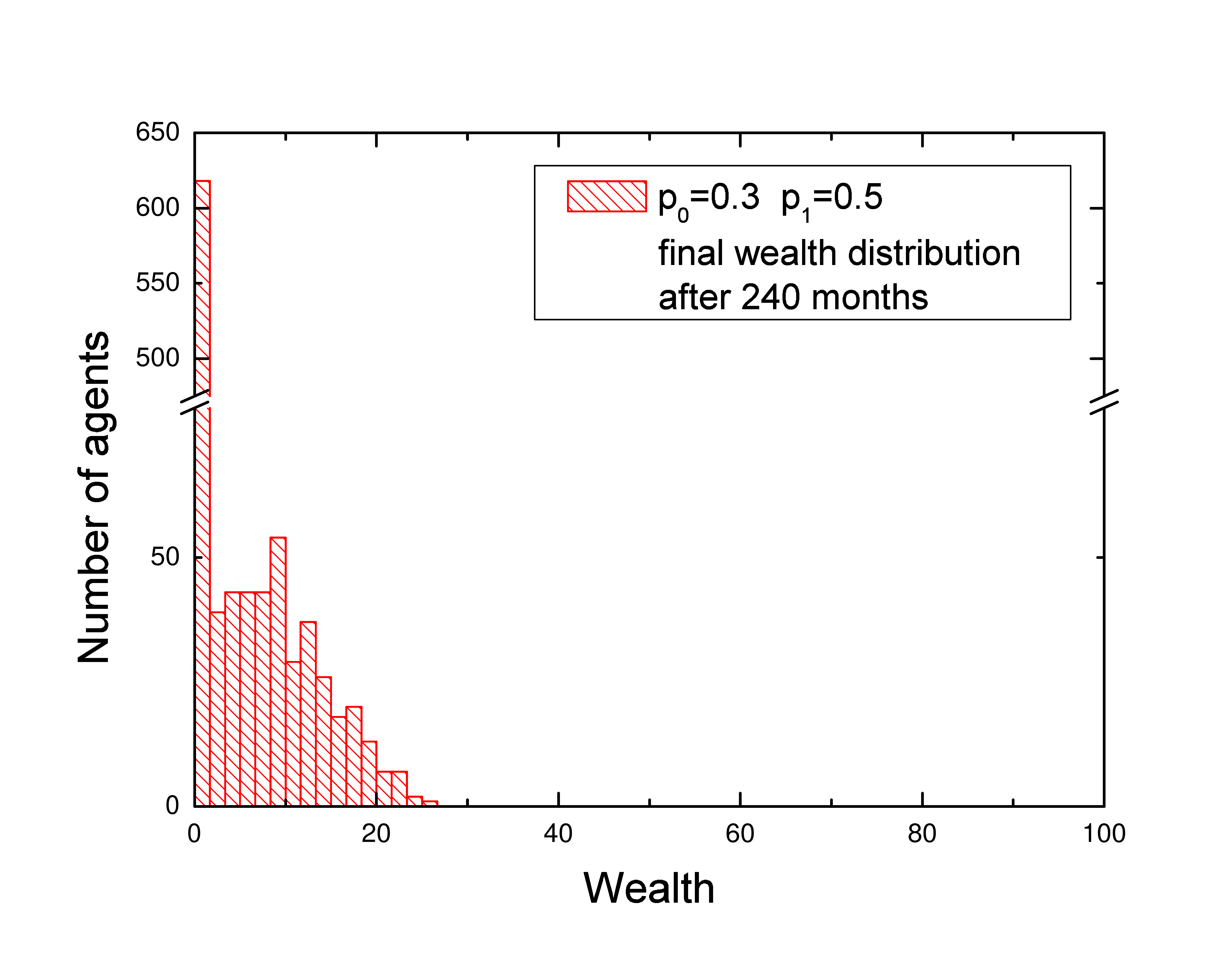}
        \includegraphics[width=8cm]{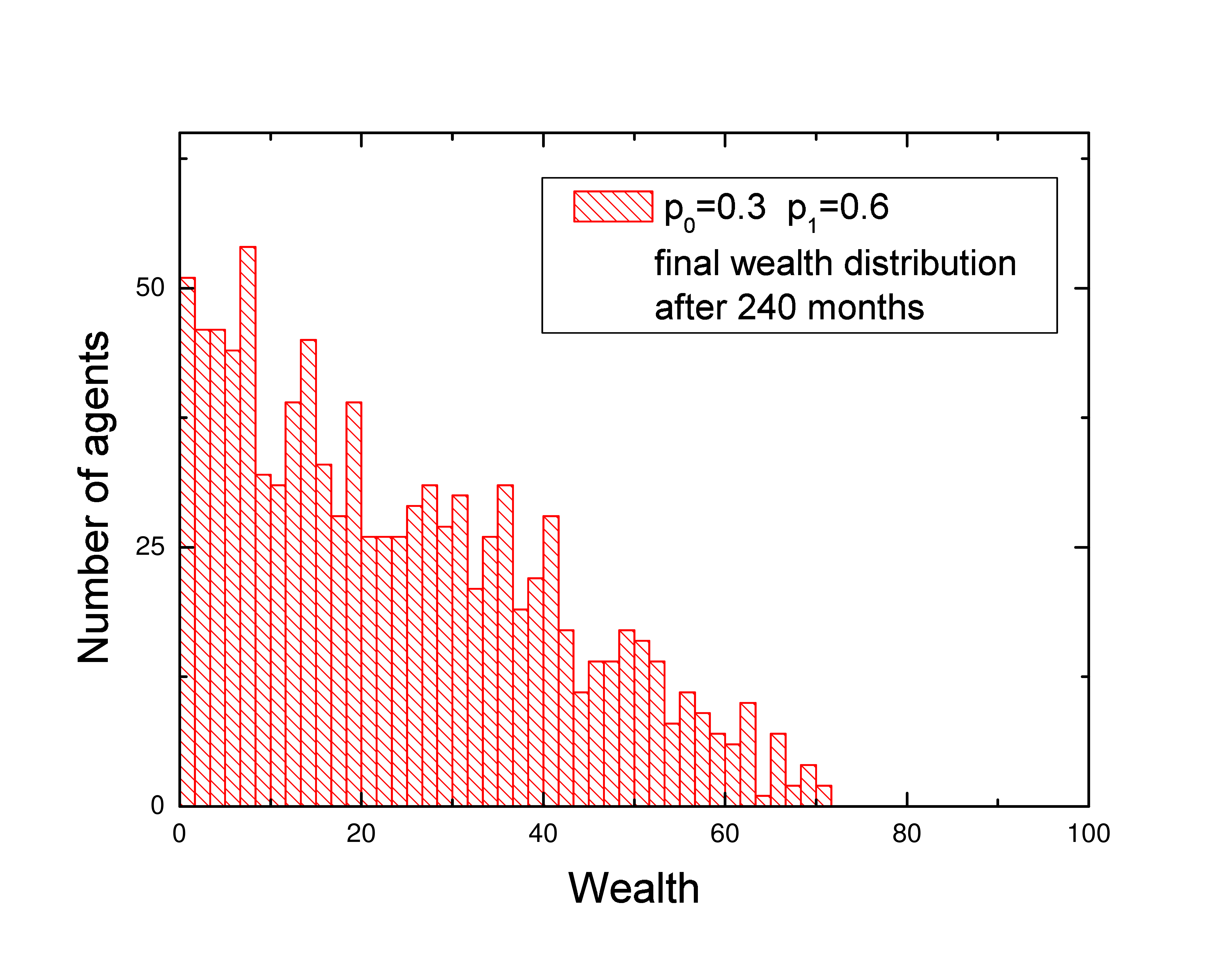}
        \includegraphics[width=8cm]{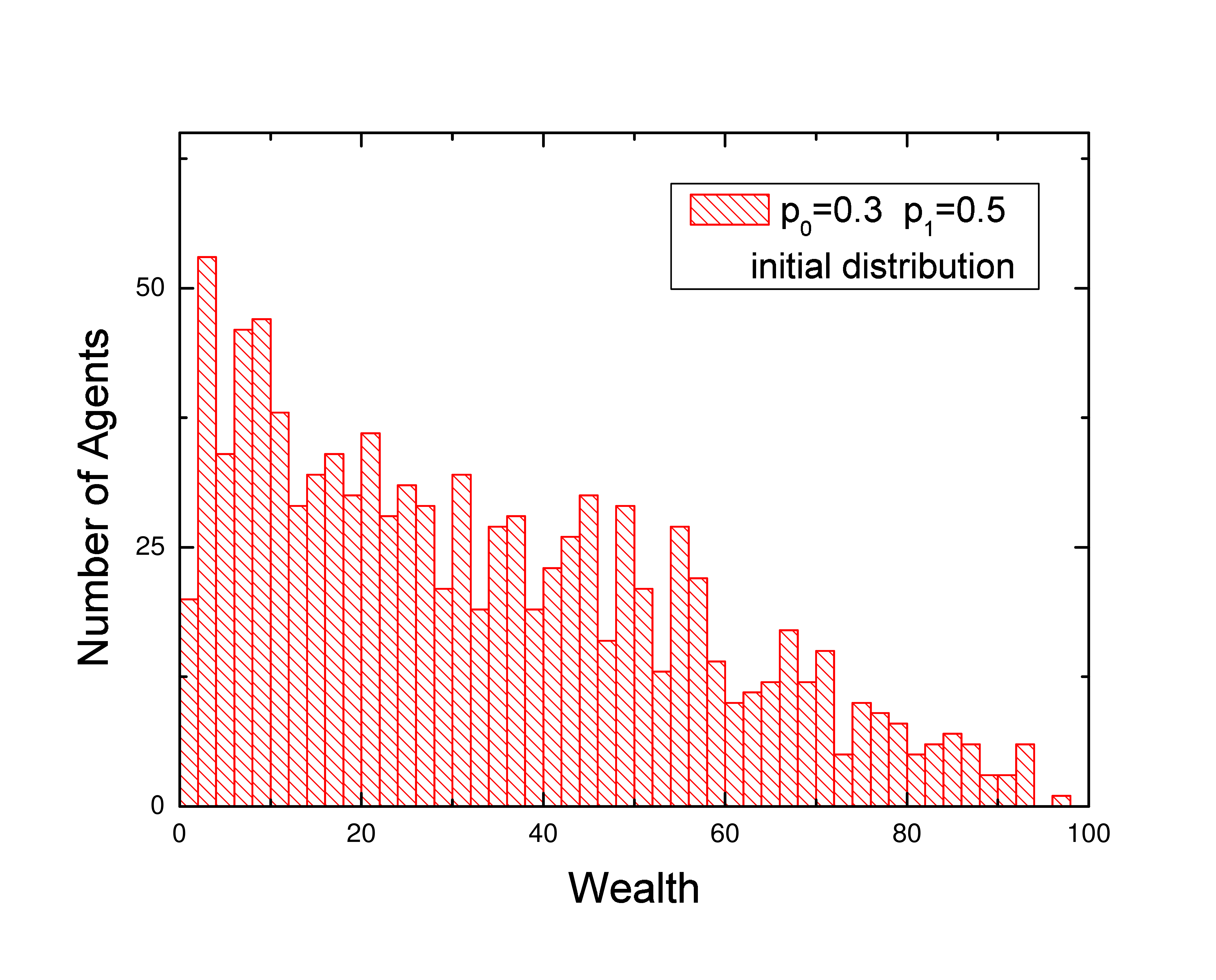}
		\end{center}
		\caption[Wealth histograms] {Histograms of the population wealth.
Upper figures: at the end of the $240$ months with $p_1=0.5$ (a)
and $p_1=0.6$ (b). Lower figure (c): initial state.}
\label{fig:WealthHistograms}
\end{figure*}

In the upper left panel of Figs.~\ref{fig:Versus_p_1} the number of
crimes and punishments per agent is represented for three representative
values of $p_0$: $0.1$, $0.3$ and $0.4$. There is a sharp transition in
the criminality when $p_1$ increases but the critical value strongly
depends on the value of $p_0$. While for $p_0=0.4$ the transition
happens for $p_1 \simeq 0.52$, for $p_0=0.1$ the critical value grows up
to $p_1=0.65$. This is an indication that the permissiveness in coping
with small crimes have a deleterious effect, since the probability of
punishment needed to deter important crimes increases.

A simple argument allows to understand the abrupt transition found in
the simulations, which is correlated with a drop in the honesty level of
the population. The average honesty level of the population increases
roughly (we neglect the influence of criminal's different dynamics) by
about $\delta H$ if crimes are punished, and decreases by the same
amount if not punished. Thus, we expect:
$\bar H(m+1) \approx \bar H(m) + (1-\pi) \delta H - \pi \delta H$ per
crime, where $\pi$ is the probability of the crime being punished.
Clearly, there should be a change from an increasing honesty dynamics to
a decresaing one for $\pi=1/2$. If crimes were punished with the same
probability whatever the value of the loot ($p_0=p_1$), the change in the
honesty dynamics would arise when this probability is equal to $1/2$. If
small crimes have less probability of punishment, $p_0<1/2$, then $p_1$
must increase to keep the same dynamics on the average.

The upper right panel shows the total capital of the society. The effect
of wrongdoing is evident. From a strictly economic point of view the
worse situation arises closely below the critical point: a high level
of criminal actions together with a relatively high frequency of
punishment (although not enough to control criminality)
have as a consequence a strong decrease in the total capital (because
the cumulated effect of booties and taxes strongly reduces the total
capital of the population). On the other hand, once the delinquency is
under control the total capital of the society arrives to a maximum level.

The opposite effect is observed in the plot of the booties and taxes
(left low panel). They are very high in the high criminality region (low
values of $p_1$) and decrease strongly when $p_1$ is above its critical
value.

Finally in the right low panel we have represented the Gini coefficient.
If we observe this figure together with the evolution of the wealth of
the society we can conclude that low criminality implies higher economic
growth and less inequality. As the Gini coefficient is an average
indicator, we present in Fig.~\ref{fig:WealthHistograms} the wealth
distribution histograms, in order to supply a complementary indicator.
The two panels on top of Fig.~\ref{fig:WealthHistograms} correspond
to the histograms for the two values of $p_0$ and $p_1$ used on figures
\ref{fig:Monthstat} and \ref{fig:MonthKapitalGini}. It is clear
that, for $p_1=0.5$, more than half of the agents have a vanishingly
small capital, so explaining the high value of the Gini coefficient,
while the total wealth of the population (represented by the total area
of the histogram) is smaller that in the case of larger $p_1$. On the
other
hand, for $p_1=0.6$ the number of agents with wealth near zero falls
down to $10\%$ of the population. Finally and just for comparison the
lower figure represents the wealth distribution for the initial state
(or, equivalently, the wages distribution).

We would like to emphasize that the results presented in this section
are averages over $100$ independent samples at the end of $240$ months
of evolution. We expect that modifying either $p_0$ or $p_1$ or both as
a function of time (as it may happen in real societies in order to
correct an abnormal increase of criminality) would produce different
results because the initial conditions before the change in the
probabilities are different (recall that here we assume a low initial
number of low honesty agents). In fact if one starts in a state of high
criminality, a very high probability of punishment (much
higher than the critical values here presented) should be needed in
order to reduce the criminality back to acceptable levels. Once more,
preventive actions should be less expensive and easier to apply than
trying to recover from a very deteriorated security situation.

\begin{figure*}
        \begin{center}
        \includegraphics[width=8cm]{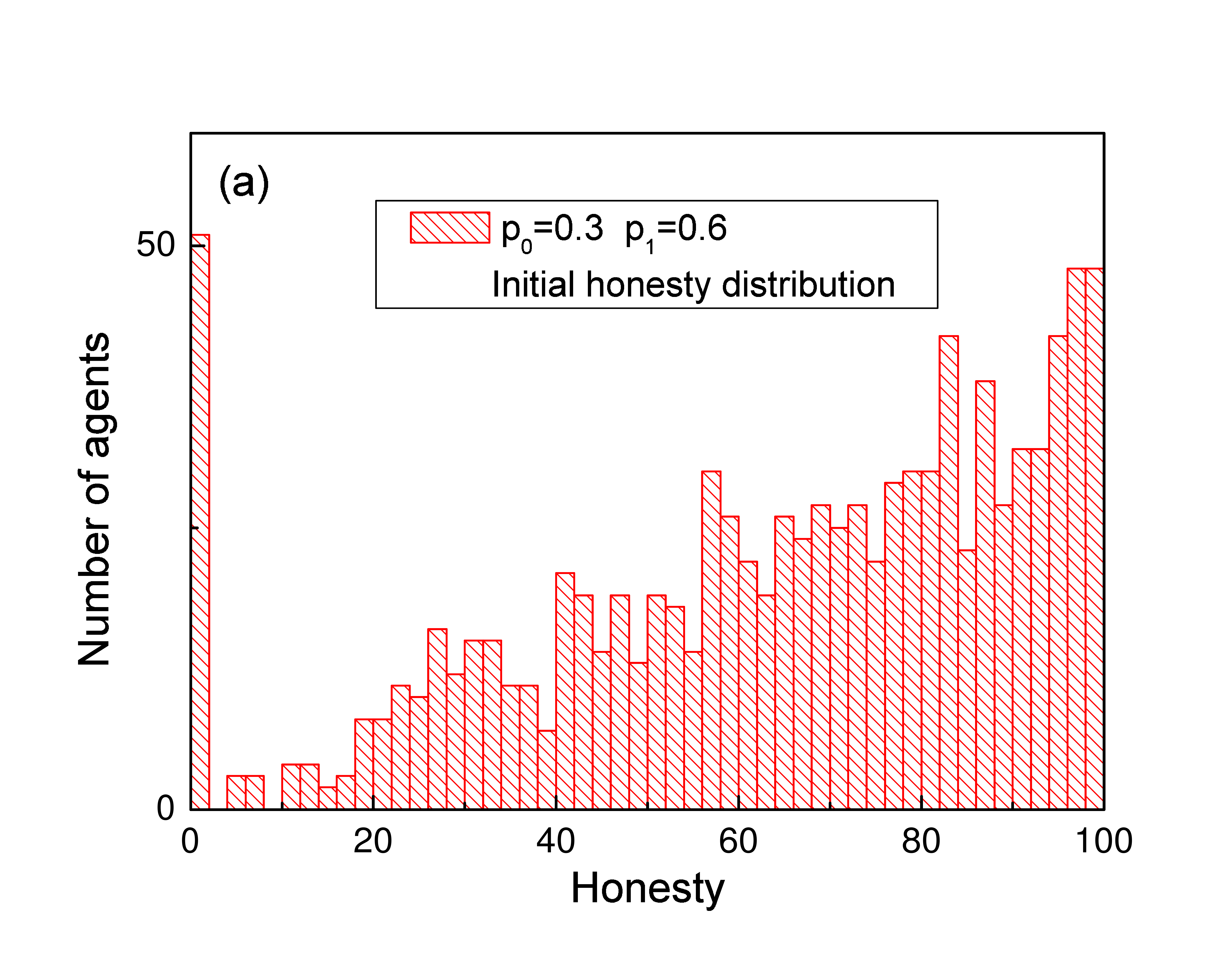}
        \includegraphics[width=8cm]{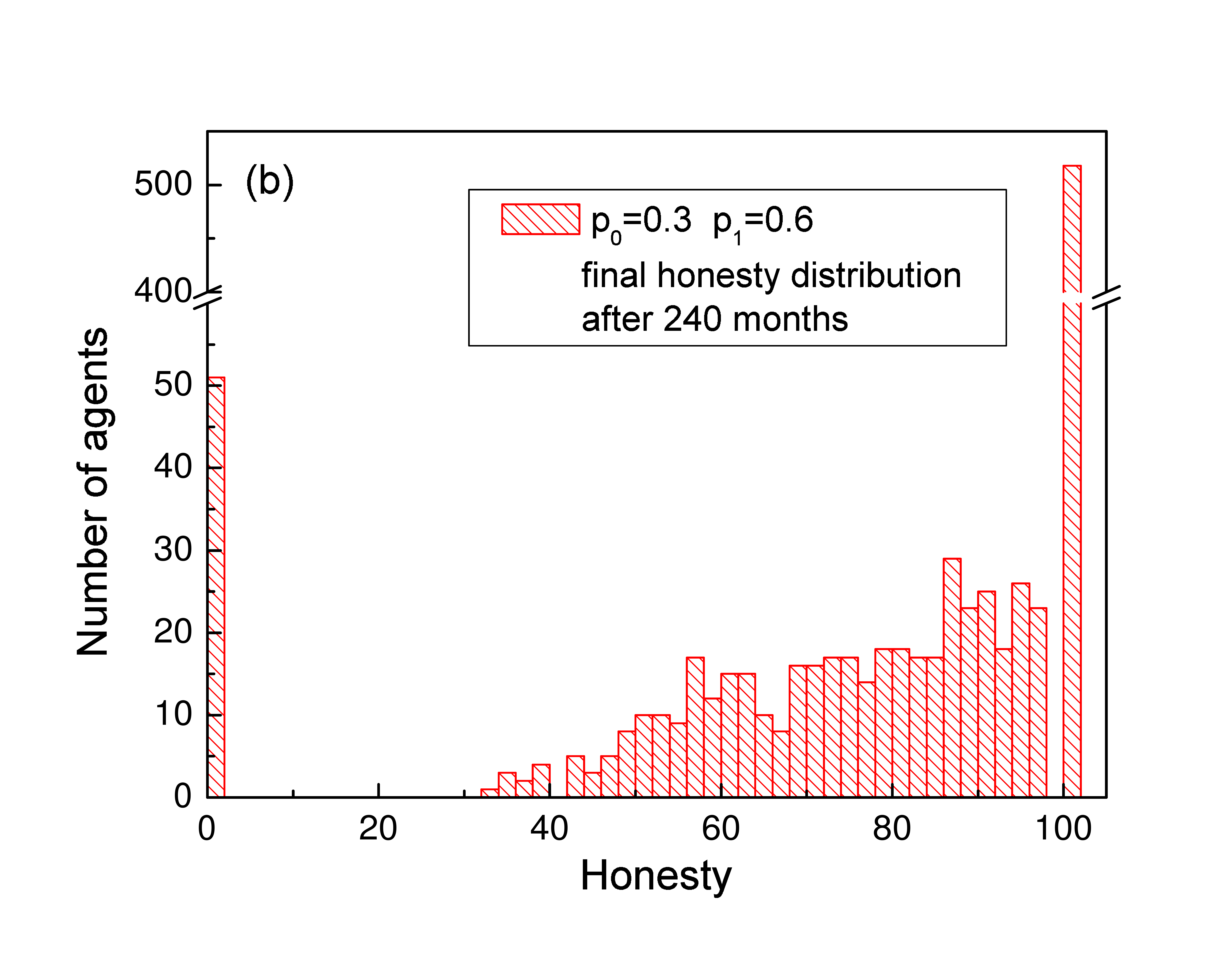}
        \end{center}
		\caption[Wealth histograms] {Histograms of the population's honesty index.
(a) initial distribution, where the small peak at zero honesty corresponds to the
initial $5 \%$ concentration of intrinsic criminals. (b) final distribution for
$p_0=0.3$ and $p_1=0.6$: more than half of the population exhibits a honesty
index $ \geq 100$ while the number of intrinsic criminals remains the initial
one. For $p_0=0.3$ and $p_1=0.5$ the honesty index of all the agents drops down
to zero (and for this reason the plot is not included): a small change in the
probability of punishment clearly induces a big change in the average honesty of
the population.}
\label{fig:honestyHistograms}
\end{figure*}

\section{Discussion and conclusion}
\label{sec:Discussion and conclusion}
A central hypothesis of our model is that the honesty level of the population is
correlated with the mere existence (or absence) of punishment, but not with its
importance (which is proportional to the size of the loot). Thus, punishing
small crimes is as effective as punishing large ones for increasing the
population's honesty globally: the honesty level increases when crimes are
punished and, on the contrary, impunity decreases it. Since small larcenies have
a lower probability of being punished than large loots, the public effort on
crime deterrence depends on the importance of the crime through the probability
of punishment.

In Fig. \ref{fig:honestyHistograms} we represented the honesty distribution for
the case $p_0=0.3$ and $p_1=0.6$ (panel (b)) and, for comparison, the initial
distribution (panel (a)). We have not represented the case $p_1=0.5$ because in
that case the honesty index of the entire population drops to zero, as discussed
above: $p_1$ is not large enough and the criminality level of the population is
the highest.

Beyond those results, our model shows an interesting abrupt drop of the
criminality level beyond a critical value of $p_1$, that depends on $p_0$,
whenever $p_0<0.5$. When small larcenies have a high probability of being
punished, the value of $p_1$ needed to reduce crime is smaller.

We remark that this behavior is very general, independent of the detailed
parameters of the simulation, as we have shown using a general argument in
section~\ref{sec:changingprob}

On the other hand, the drop in criminality has very
positive consequences: increase of the global earnings because taxes
decrease, and stabilization of the inequality at the level
corresponding to the differences in wages.

So, to conclude, we have presented a simple model of crime and punishment
that stands on the assumption that punishment has a deterrent effect on
criminality. Our main result is that tolerance with respect to small felonies
(small value of $p_0$) has a global negative consequence because it requires
bigger efforts to cope with important crimes in order to keep a given level of
honesty. We also observe an avalanche effect since a small change in the
probability of punishment may reduce or increase the average criminality in a
very significative way. Also, the economic consequences of criminality are
remarkable, both in the observed total wealth of the population, as well as in
the measure of wealth inequality.

A less crude model should also include the effect of a particular treatment
(or not) of recidivism. This is a point of discussion in countries like France,
where some legislators ask for a minimum sentence for relapse. Also, one should
be more careful in treating more ``sophisticated'' criminality, like organized
crime, or criminals that choose the victim according to the expected loot. It
would be interesting to analyze the effect of imprisonment: either to recover or
to increase the inmates criminal tendencies. We are presently working on these
points, which are of great importance.

\bibliographystyle{unsrt}
\bibliography{CC_gordon_2007}

\begin{thebibliography}{10}

\bibitem{genesis}
Bible.
\newblock {\em The Holy Bible, The Book of Genesis}.
\newblock New York: American Bible Society: 1999, 4,8, king {J}ames {V}ersion
  edition, (1999).

\bibitem{genesis2}
Bible.
\newblock {\em The Holy Bible, The Book of Genesis}.
\newblock New York: American Bible Society: 1999, 27,1, king {J}ames {V}ersion
  edition, (1999).

\bibitem{VolterraConsulting03}
Robin~Marris et~al.
\newblock Modelling crime and offending: recent developments in {E}ngland and
  {W}ales.
\newblock Technical report, Home Office, RDS website {\em
  http://www.homeoffice.gov.uk/rds/pdfs2/ occ80modelling.pdf}, (2003).

\bibitem{Salas06}
Denis Salas.
\newblock Punir n'est pas la seule finalit\'e.
\newblock {\em L'{E}xpress Sep 6, 2006 {\em http://www.lexpress.fr/info
  /quotidien/actu.asp?id=5964}}, (2006).

\bibitem{argent}
Constitutional Convention.
\newblock {\em Constituci\'on de la {N}aci\'on {A}rgentina}.
\newblock Primera parte, Cap\'itulo Primero, Art. 18, (1994).

\bibitem{Blumstein02}
A.~Blumstein.
\newblock Crime modeling.
\newblock {\em Operations Research}, {\bf 50}/1:16--24, (2002).

\bibitem{Becker68}
G.~Becker.
\newblock Crime and punishment: an economic approach.
\newblock {\em Journal of Political Economy}, {\bf 76}:169--217, (1968).

\bibitem{Ehrlich75}
I.~Ehrlich.
\newblock The deterrent effect of capital punishment: A question of life and
  death.
\newblock {\em American Economic Review}, {\bf 65}:397--417, (1975).

\bibitem{Ehrlich96}
I.~Ehrlich.
\newblock Crime, punishment, and market for offenses.
\newblock {\em The Journal of Political Perspectives}, {\bf 10}:43--67, (1996).

\bibitem{DeutschSpiegelTempleman92}
Joseph Deutsch, Uriel Spiegel, and Joseph Templeman.
\newblock Crime and income inequality: An economic approach.
\newblock {\em AEJ}, {\bf 20}:46--54, (1992).

\bibitem{BourguignonNunezSanchez02}
F.~Bourguignon, J.~Nunez, and F.~Sanchez.
\newblock What part of the income distribution does matter for explaining
  crime? the case of colombia.
\newblock {\em Working paper N° 2003-04, DELTA}, {\em http://www.delta.ens.fr},
  (2002).

\bibitem{DahlbergGustavsson05}
Matz Dahlberg and Magnus Gustavsson.
\newblock Inequality and crime: separating the effects of permanent and
  transitory income.
\newblock {\em Working paper 2005:19, Institute for Labour Market Policy
  Evaluation (IFAU)}, (2005).

\bibitem{Levitt99}
Steven~D. Levitt.
\newblock The changing relationship between income and crime victimization.
\newblock {\em FRBNY ECONOMIC POLICY REVIEW / SEPTEMBER 1999}, {\bf }:88--98,
  (1999).

\bibitem{Levitt97}
S.~Levitt.
\newblock Using electoral cycles in police hiring to estimate the effect of
  police on crime.
\newblock {\em American Economic Review}, {\bf 87}(3):270--290, (1997).

\bibitem{FreemanRB96}
Richard~B. Freeman.
\newblock Why do so many young american men commit crimes and what might we do
  about it?
\newblock {\em Journal of Economic Perspectives}, {\bf 10}:25--42, (1996).

\bibitem{Eide99}
Erling Eide.
\newblock {\em Economics of Criminal Behavior}, chapter 8100, pages 345--389.
\newblock Edward Elgar and University of Ghent, {\em
  http://encyclo.findlaw.com/8100book.pdf}, (1999).

\bibitem{VolterraMarris00}
Robin Marris and Volterra Consulting.
\newblock Survey of the research litterature on the criminological and economic
  factors influencing crime trends.
\newblock Technical report, HomeOffice, RDS website
  (http://www.homeoffice.gov.uk/rds/pdfs2/ occ80modellingsup.pdf), (2003).

\bibitem{NaPhGoVa05}
J.-P. Nadal, D.~Phan, M.~B. Gordon, and J.~Vannimenus.
\newblock Multiple equilibria in a monopoly market with heterogeneous agents
  and externalities.
\newblock {\em Quantitative Finance}, {\bf 5}(6):557--568, (2006).
\newblock Presented at the 8th Annual Workshop on Economics with Heterogeneous
  Interacting Agents (WEHIA 2003).

\bibitem{GoNaPhSe07}
M.~B. Gordon, J.-P. Nadal, D.~Phan, and V.~Semeshenko.
\newblock Discrete choices under social influence: generic properties.
\newblock {\em {\em http://halshs.archives-ouvertes.fr/halshs-00135405}},
  (2007).

\bibitem{GlaeserSacerdoteScheinkman96}
E.~L. Glaeser, B.~Sacerdote, and J.~A. Scheinkman.
\newblock Crime and social interactions.
\newblock {\em Quarterly Journal of Economics}, {\bf 111}:507--548, (1996).

\bibitem{CampbellOrmerod00}
M.~Campbell and P.~Ormerod.
\newblock Social interactions and the dynamics of crime.
\newblock {\em Volterra Consulting Preprint}, {\bf }, (1997).

\bibitem{web_dando}
Jill Dando~Institute of~Crime~Science, (2001).

\end{thebibliography}

\end{document}